\def\sgn{{\text{sgn\,}}}
\def\be{\begin{equation}}
\def\ee{\end{equation}}
\def\bea{\begin{eqnarray}}
\def\eea{\end{eqnarray}}
\def\bse{\begin{subequations}}
\def\ese{\end{subequations}}

\documentclass[prl,twocolumn,showpacs,preprintnumbers,amsmath,amssymb]{revtex4}
\usepackage{graphicx}% Include figure files
\usepackage{dcolumn}% Align table columns on decimal point
\usepackage{bm}% bold math
\begin{document}
\preprint{NSF-KITP-07-211}
\title{Analogy between Three-Dimensional Helimagnetic Metals and
                     Two-Dimensional Nonmagnetic Metals: Transport in the Weak-Disorder
                     Regime%\\
\vskip 1mm
       %\small{\rm{(KITP Preprint NSF-KITP-05-31)}}
       %\small{$[$ Phys. Rev. B {\bm 63}, 174428 (2001) $]$
}
\author{T.R. Kirkpatrick$^{1}$, D. Belitz$^{2,3}$, and Ronojoy Saha$^{2}$}
\affiliation{$^{1}$Institute for Physical Science and Technology and Department
                   of Physics, University of Maryland, College Park, MD 20742\\
             $^{2}$Department of Physics and Institute of Theoretical Science,
                   University of Oregon, Eugene, OR 97403\\
             $^{3}$Kavli Institute for Theoretical Physics, University of California,
                   Santa Barbara, CA 93106}
\date{\today}
\begin{abstract}
We present a quasi-particle model that allows for a simple description of the
electronic properties of metallic helimagnets. For weak quenched disorder, we
find a leading linear temperature dependence of the electrical conductivity for
3-$d$ materials. This is reminiscent of the behavior of nonmagnetic 2-$d$
systems, and reflects a general tendency of certain properties of bulk
helimagnets to appear effectively 2-$d$. The sign of the effect is opposite to
that in nonmagnetic 2-$d$ materials. These surprising predictions should be
observable in weak helimagnets.
%
% 559 characters
\end{abstract}

\pacs{75.30.Ds; 75.30.-m; 75.50.-y; 75.25.+z}

\maketitle

Helimagnets (e.g., MnSi, or FeGe) are magnetic materials where the
magnetization shows ferromagnetic order in any plane perpendicular to a certain
axis, but points in different directions depending on the position along that
axis, forming a spiral. The metallic helimagnet MnSi is a well-studied material
with unusual properties. It shows helical order below a critical temperature
$T_{\text{c}} \approx 30\,{\text K}$ with a helix wavelength $2\pi/q \approx
180\,\AA$ \cite{Ishikawa_et_al_1976}. Hydrostatic pressure decreases
$T_{\text{c}}$ until the long-range order \cite{LRO_footnote} disappears at a
critical pressure $p_{\,\text{c}} \approx 14\,{\text{kbar}}$
\cite{Pfleiderer_et_al_1997}. At higher pressure, there is a phase or region
where short-range helical order persists and the electrical resistivity $\rho$
shows a $T^{3/2}$ temperature dependence in a range between a few mK and
several K \cite{Pfleiderer_Julian_Lonzarich_2001}. If this were the true
asymptotic low-$T$ behavior, it would represent non-Fermi liquid behavior in a
bulk material, which would be very remarkable.

The samples used in Ref.\ \onlinecite{Pfleiderer_Julian_Lonzarich_2001} have a
residual resistivity of about $0.3\,\mu\Omega{\text{cm}}$, which corresponds to
a rather large elastic mean-free time $\tau$. At the experimentally attainable
temperatures, this places them in the weak-disorder regime, $T\tau \gg 1$
\cite{ballistic_footnote}, where the transport is governed by free-electron
motion in between rare scattering events, as opposed to diffusive motion in the
opposite limit, $T\tau \ll 1$. The weak-disorder regime has been investigated
by Zala et al. \cite{Zala_Narozhny_Aleiner_2001} for the case of electrons
interacting via a screened Coulomb interaction in non-magnetic 2-$d$ metals,
where they found a linear $T$ dependence of the resistivity. (In 3-$d$, the
corresponding resistivity correction is of $O(T^2 \ln T)$
\cite{Sergeev_Reizer_Mitin_2005}.) Although the weak-disorder regime is never
the true asymptotic low-temperature regime, it thus can display very unusual
behavior. Furthermore, depending on the impurity concentration, it can
represent the low-$T$ asymptotics for practical purposes.

In a helimagnet, massless fluctuations of the helix (helimagnons) that couple
to the electrical conductivity are an obvious possible source for unusual
transport behavior, and it is natural to first study their effects in the
ordered phase, before trying to understand their potential ramifications more
generally \cite{pinning_footnote}. It was recently shown that helimagnons do
indeed lead to a non-analytic temperature dependence of the resistivity in the
ordered phase \cite{Belitz_Kirkpatrick_Rosch_2006b}. However, for the most
interesting observables in a clean system these effects provide {\em
corrections} to the usual Fermi-liquid behavior: the leading helimagnon
contribution to the resistivity has a $T^{5/2}$ behavior, and the specific heat
goes as $T^2$, although the single-particle relaxation rate goes as $T^{3/2}$.

In this Rapid Communication we investigate the resistivity of 3-$d$ helimagnets in the
weak-disorder regime (which is defined slightly differently than in the Coulomb
case, see below). We show that the leading temperature correction $\delta\rho
(T)$ to the resistivity is proportional to $T$ due to scattering by
helimagnons. Remarkably, this is much stronger than either the $T^{5/2}$ clean
helimagnon contribution or the $T^2$ Fermi-liquid contribution, and it is
reminiscent of the behavior of 2-$d$ nonmagnetic metals. In contrast to the
latter, however, the sign of the $T$-dependence is antilocalizing: there is a
$T$-independent part that decreases the resistivity, and the $T$-dependent part
is $\delta\rho \propto +T\tau$.

Transport theory for helimagnets is rather complicated, even in the clean case,
if the Kubo formula is evaluated using the usual plane-wave basis. In this
basis, the electronic Green function is not diagonal in either wave vector
space or spin space, which makes for very cumbersome calculations
\cite{Belitz_Kirkpatrick_Rosch_2006b}. Including impurity scattering in this
formalism would be hard. However, several features of this theory, which we
quote below, suggest a much simpler effective description. The poles of the
electronic Green function, or quasi-particle (QP) energies, are given by
\be
\omega_{1,2}({\bm k}) = \frac{1}{2} \left(\xi_{\bm k} + \xi_{{\bm k}+{\bm q}}
\pm \sqrt{(\xi_{\bm k} - \xi_{{\bm k}+{\bm q}})^2 + 4\lambda^2}\right).
\label{eq:1}
\ee
Here $\xi_{\bm k} = \epsilon_{\bm k} - \epsilon_{\text{F}}$, with
$\epsilon_{\bm k}$ the electronic energy-momentum relation and
$\epsilon_{\text{F}}$ the chemical potential. For a cubic crystal, such as
MnSi,
\be
\epsilon_{\bm k} = \frac{{\bm k}^2}{2m_{\text e}} + \frac{\nu}{2m_{\text e}
k_{\text{F}}^2} \left(k_x^2 k_y^2 + k_y^2 k_z^2 + k_z^2 k_x^2\right) + O(k^6),
\label{eq:2}
\ee
with $m_{\text e}$ the effective electron mass, and $\nu = O(1)$ a
dimensionless measure of deviations from nearly free electrons. We choose units
such that $\hbar = k_{\text{B}} = e = 1$. ${\bm q} = (0,0,q)$ is the helix
pitch vector, which we choose to point in the $z$-direction, and $\lambda$ is
the Stoner splitting. The two signs of the square root represent the two Stoner
bands, and for ${\bm q}=0$ one recovers the usual ferromagnetic result in
Stoner approximation. Excitations between the two Stoner bands will be gapped
by $\lambda$, and we can thus restrict ourselves to a single band by
considering spinless QPs with a resonance frequency $\omega_1({\bm k})$.

For the electronic QP Green function one thus expects
\be
G_0(p) = 1/(i\omega_n - \omega_{1}({\bm p})),
\label{eq:3}
\ee
where $p = ({\bm p},i\omega_n)$ with $\omega_n$ a fermionic Matsubara frequency
\cite{diagonality_footnote}. We have shown that such a QP Green function can
indeed be derived from a canonical transformation to suitable fermionic degrees
of freedom $\eta(p)$ \cite{us_tbp}. The soft, or massless, helical degrees of
freedom, the helimagnons, are fluctuations of a generalized helical phase
$\phi$ with a frequency
\bse
\label{eqs:4}
\be
\omega_0({\bm p}) = \sqrt{c_z p_z^{\,2} + c_{\perp}p_{\perp}^{\,4}}.
\label{eq:4a}
\ee
Here $p_z$ and $p_{\perp}$ are the longitudinal and transverse components,
respectively, of the wave vector with respect to the pitch wave vector ${\bm
q}$, and the elastic constants $c_z$ and $c_{\perp}$ are given by
\be
c_z = \gamma_z \lambda^2\,q^2/k_{\text{F}}^4\quad,\quad c_{\perp} =
\gamma_{\perp} \lambda^2/k_{\text{F}}^4.
\label{eq:4b}
\ee
\ese
$\gamma_z$ and $\gamma_{\perp}$ are model-dependent numbers. For the model
considered in Ref.\ \cite{Belitz_Kirkpatrick_Rosch_2006b}, their values are
$\gamma_z = 1/36$ and $\gamma_{\perp} = 1/96$. Equation (\ref{eq:4a}) is valid
for $\vert{\bm p}\vert < q$. Notice that the helimagnon dispersion relation is
anisotropic: it is ferromagnet-like in the longitudinal direction, but
antiferromagnet-like in the transverse direction. The helimagnon susceptibility
is given by
\be
\chi(k) = \langle \phi(k)\,\phi(-k)\rangle =
\frac{1}{2N_{\text{F}}}\,\frac{q^2/3k_{\text{F}}^2}{\omega_0^2({\bm k}) -
(i\Omega_n)^2}\ ,
\label{eq:5}
\ee
with $\Omega_n$ a bosonic Matsubara frequency, $k = ({\bm k},i\Omega_n)$, and
$N_{\text{F}}$ the density of states at the Fermi level. We see that frequency
or temperature scale with the soft wave vector ${\bm k}$ as $T \sim k_z \sim
k_{\perp}^2$.

\begin{figure}[t]
\vskip -0mm
\includegraphics[width=7.0cm]{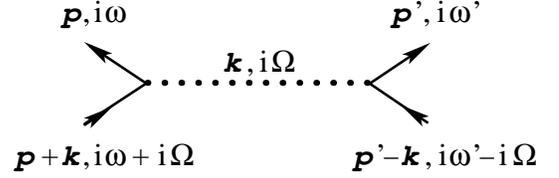}
\caption{Effective interaction between QPs (solid lines) by
 exchange of a helimagnon (dotted line).}
\label{fig:1}
\end{figure}

Since $\phi$ is a phase, only the gradient of $\phi$ is of physical
significance and will couple to the QPs. Furthermore, the results of Ref.\
\onlinecite{Belitz_Kirkpatrick_Rosch_2006b} show that the leading coupling is
not to the QP density, but rather to the $z$-$\perp$ components of the QP
stress. Upon integrating out $\phi$, one thus expects an effective potential
for the interaction of QPs by means of exchange of helimagnons that is given
pictorially in Fig.\ \ref{fig:1},
and analytically by
\bse
\label{eqs:6}
\be
V_{\text{eff}} = -\lambda^2 (q^2/8 m_{\text{e}}^2)\,\chi(k)\,\gamma({\bm
k},{\bm p})\,\gamma({\bm k},{\bm p'}),
\label{eq:6a}
\ee
with a vertex function
\be
\gamma({\bm k},{\bm p}) = \nu\,p_z ({\bm p}_{\perp}\cdot{\bm
k}_{\perp})/\lambda\,k_{\text{F}}^2.
\label{eq:6b}
\ee
\ese

Finally, we expect static impurities or quenched disorder to couple to the QP
density. Putting everything together, we now have the following effective
action:
\bse
\label{eqs:7}
\be
S = S_0 + S_{\text{int}} + S_{\text{dis}},
\label{eq:7a}
\ee
where
\be
S_0 = (T/V)\sum_p\ {\bar\eta}(p) \left[i\omega_n - \omega_{1}(p)\right] \eta(p)
\label{eq:7b}
\ee
describes free QPs,
\bea
S_{\text{int}} &=& \frac{-T}{V} \sum_{k} \frac{1}{V^2}\sum_{{\bm p},{\bm p}'}
V_{\text{eff}}(k;{\bm p},{\bm p}')\ T\sum_{i\omega} {\bar\eta}({\bm p},i\omega)
\nonumber\\
&& \hskip -40pt \times \eta({\bm p}+{\bm k},i\omega + i\Omega)\ {\bar\eta}({\bm
p}',i\omega)\,\eta({\bm p}'-{\bm k},i\omega - i\Omega)
\label{eq:7c}
\eea
describes the effective QP interaction via the exchange of a helimagnon, and
\be
S_{\text{dis}} = (-1/V^2)\sum_{{\bm k},{\bm p}} u({\bm k}-{\bm
p})\,T\sum_{i\omega} {\bar\eta}({\bm k},i\omega)\, \eta({\bm p},i\omega)
\label{eq:7d}
\ee
\ese
describes the quenched disorder. Here $u$ is a delta-correlated point-like
random potential with a Gaussian distribution whose second moment is given by
$1/2\pi N_{\text{F}}\tau$.

The Eqs.\ (\ref{eqs:7}) represent an effective model for QPs interacting via a
dynamical potential that allows for an evaluation of the Kubo formula in
complete analogy to electrons interacting via a dynamically screened Coulomb
interaction. The only unusual aspects are, (1) the coupling to the QP stress
rather than the density, and (2) the anisotropic nature of the exchanged
excitations. The former is embodied in the vertex functions $\gamma$ in Eqs.\
(\ref{eqs:6}) and easy to handle technically, but has important physical
consequences. The latter makes the bulk system behave in some respects like a
$2$-$d$ system. We have derived this effective model from the one studied in
Ref.\ \onlinecite{Belitz_Kirkpatrick_Rosch_2006b} by means of a canonical
transformation \cite{us_tbp}; the above considerations represent just
plausibility arguments.

Transport theory now proceeds in analogy to the Coulomb problem. In order to
calculate the static electrical conductivity tensor $\sigma_{ij}$ we evaluate
the Kubo formula
\bea
\sigma_{ij} &=& -{\text{lim}}_{\,\Omega\rightarrow 0}\,
{\text{Re}}\,\frac{i}{i\Omega_n} \int_0^{1/T} d\tau\ e^{i\Omega_n\tau}\
\left\langle T_{\tau}\, {\hat j}_i({\bm k},\tau)\right.
\nonumber\\
&&\hskip 30pt \times \left. {\hat j}_j({\bm k},\tau=0)\right\rangle
\big\vert_{i\Omega_n \rightarrow \Omega + i0,\ {\bm k}=0}\ ,
\label{eq:8}
\eea
with ${\hat{\bm j}}({\bm k},\tau)$ the current operator in imaginary-time
representation and $T_{\tau}$ the imaginary-time ordering operator. The tensor
$\sigma_{ij}$ is diagonal and has two independent elements, $\sigma_{\text{L}}
\equiv \sigma_{zz}$ and $\sigma_{\perp} \equiv \sigma_{xx} = \sigma_{yy}$.

In the clean limit, $\tau \rightarrow \infty$, we reproduce the results of
Ref.\ \onlinecite{Belitz_Kirkpatrick_Rosch_2006b}: an infinite ladder
resummation of both the self-energy and vertex-correction contributions to the
conductivity shows that the leading low-$T$ terms cancel between the two sets
of diagrams. The transport relaxation rate, and hence the resistivity, shows a
$T^{5/2}$ behavior, whereas the single-particle relaxation rate goes as
$T^{3/2}$.

\begin{figure}[b]
\vskip -0mm
\includegraphics[width=8.0cm]{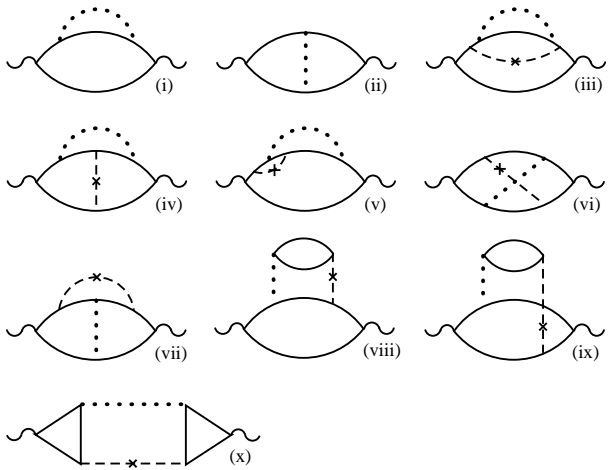}
\caption{Leading contributions to the conductivity in the weak-disorder limit.
 Solid lines represent the Green function given by Eq.\ (\ref{eq:9}), dotted
 lines represent the effective potential, Eqs.\ (\ref{eqs:6}), and dashed
 lines represent the impurity factor $1/2\pi N_{\text{F}}\tau$. Obvious
 symmetric counterparts of all diagrams except (ii) are not shown. See
 the text for additional information.}
\label{fig:2}
\end{figure}

In the weak-disorder limit, it is convenient to include the disorder in the
Green function in the Born approximation, i.e., to use a Green function
\be
G(p) = 1/(i\omega_n - \omega_1({\bm p}) + i\,\sgn(\omega_n)/2\tau)
\label{eq:9}
\ee
instead of Eq.\ (\ref{eq:3}). The diagrams that contribute to the correction to
the conductivity to leading order in the disorder in the weak-disorder limit
are shown in Fig.\ \ref{fig:2}; they are the same as in Ref.\
\onlinecite{Zala_Narozhny_Aleiner_2001}.

Diagrams (i) and (ii) also contribute in the clean limit, where their
respective ladder resummations must be taken into account. In the presence of
disorder, their leading terms are proportional to $\tau^2 T^{3/2}$. (This is
the result of adding the clean single-particle rate, which is proportional to
$T^{3/2}$, to $1/\tau$ according to Matthiessen's rule, and expanding.) These
terms cancel between diagrams (i) and (ii), as they must according to Ref.\
\onlinecite{Belitz_Kirkpatrick_Rosch_2006b}. The calculation shows that the
small parameter for the disorder expansion is $\epsilon = 1/\sqrt{\tau^2 T
\epsilon_{\text{F}}^2/\lambda}$. This is different from the Coulomb case, where
the small parameter is $1/\tau T$ \cite{Zala_Narozhny_Aleiner_2001}.
Consequently, the next-leading terms are of order $\tau T$. These also cancel
between the two diagrams.

For the remaining diagrams it suffices to calculate their leading behavior, and
they can be classified with respect to their momentum structure. They each
contain six Green functions that factorize into two correlation functions
formed by $n$ and $6-n$ Green functions, respectively, with $n=3$ or $n=4$.
Power counting shows that the $(4,2)$ partitions are smaller than the $(3,3)$
partitions by a factor of $(q/k_{\text{F}})^2$. Dropping the former leaves us
with diagrams (iii), (iv), (vi). Diagram (x) also is a $(3,3)$ partition, but
its leading contribution vanishes due to reality requirements. Furthermore,
power counting reveals that for $\sigma_{\perp}$, only diagram (iii) is of
$O(\tau T)$, whereas for $\sigma_{\text{L}}$ one needs to consider diagrams
(iv) and (vi) as well.

We thus conclude that the leading contributions to both the longitudinal and
the transverse conductivity corrections are of $O(\tau T)$. We find
\be
\delta\sigma_{\text{L}} = 3\delta\sigma_{\perp} =
-\sigma_0\,\frac{\pi\nu^2}{192\sqrt{\gamma_z\gamma_{\perp}}}\,
\left(\frac{\epsilon_{\text{F}}}{\lambda}\right)^2\,
\left(\frac{q}{k_{\text{F}}}\right)^3\,\frac{T}{\epsilon_{\text{F}}}\ ,
\label{eq:10}
\ee
with $\sigma_0 = (2k_{\text{F}}/3\pi^2)\,\epsilon_{\text{F}}\tau$ the Drude
conductivity. In writing this result we have dropped a $T$-independent
UV-cutoff dependent contribution to $\delta\sigma$ whose sign is opposite that
of the $T$-dependent contribution.

We now discuss this result. First, notice the sign of the result.
$\delta\sigma$ increases with decreasing $T$, which means the effect is
antilocalizing. This is in contrast to the Coulomb case, where the sign is
localizing. The sign is the same for both the self-energy diagram (iii), and
the vertex correction diagrams (iv) and (vi). In the former, it is a result of
the angular dependence of the effective interaction, Eq.\ (\ref{eq:6b}), which
in turn is a result of the stress coupling, and which suppresses backscattering
in diagram (iii). For a density coupling, diagram (iii) would be localizing.

Second, we discuss the range of validity of our result. The weak-disorder
temperature regime is bounded below by the requirement that the expansion
parameter $\epsilon$ be small, i.e., $T\tau^2 \epsilon_{\text{F}}^2/\lambda \gg
1$. This defines a temperature scale $T^* =
\lambda/(\epsilon_{\text{F}}\tau)^2$. It is bounded above by the weak-disorder
correction to the relaxation rate crossing over to either the clean-limit rate
$1/\tau_{\text{clean}} \propto T^{5/2}$ \cite{Belitz_Kirkpatrick_Rosch_2006b},
or to the Fermi-liquid rate $1/\tau_{\text{FL}} \approx
T^2/\epsilon_{\text{F}}$. For the former, we have a crossover temperature
$T_{1-5/2} = \lambda/(\epsilon_{\text{F}}\tau)^{2/3}$, for the latter, $T_{1-2}
= (1/\tau)(q/k_{\text{F}})^3(\epsilon_{\text{F}}/\lambda)^2$. The weak-disorder
rate thus yields the dominant $T$ dependence of the resistivity in the regime
\be
T^* \ll T \ll \text{Min}(T_{1-2},T_{1-5/2}),
\label{eq:11}
\ee
and $T_{1-2}/T_{1-5/2} = (q/k_{\text{F}})^3(\epsilon_{\text{F}}/\lambda)^3
/(\epsilon_{\text{F}}\tau)^{1/3}$. If $\lambda \alt \epsilon_{\text{F}}$, then
$T_{1-2} \ll T_{1-5/2}$. In a weak helimagnet, where
$\lambda/\epsilon_{\text{F}}$ might be as small as $q/k_{\text{F}}$, we still
have $T_{1-2} < T_{1-5/2}$ on account of the factor
$1/(\epsilon_{\text{F}}\tau)^{1/3}$. For realistic parameter values, the
weak-disorder regime will thus be
\be
T^* \ll T \ll T_{1-2} = T^*\,(\epsilon_{\text{F}}\tau)
(q/k_{\text{F}})^3(\epsilon_{\text{F}}/\lambda)^3.
\label{eq:12}
\ee
The upper limit depends strongly on $\epsilon_{\text{F}}/\lambda$, which is
often not well known. For weak helimagnets, with $\lambda \ll
\epsilon_{\text{F}}$, it is possible to have $T_{1-2} \gg T^*$ by virtue of
$\epsilon_{\text{F}}\tau \gg 1$. For larger values of $\lambda$ the
weak-disorder correction may never dominate unless the system is extremely
clean. In that case, one will have to subtract the Fermi-liquid $T^2$ behavior
%from the resistivity
in order to observe the weak-disorder contribution. For MnSi, with
$\epsilon_{\text{F}} \approx 23,000\,{\text{K}}$ and $\epsilon_{\text{F}}\tau
\approx 1,000$ for the samples of Ref.\
\onlinecite{Pfleiderer_Julian_Lonzarich_2001}, one has $T^* \approx
10\,\text{mK}$. $q/k_{\text{F}} \approx 0.02$,
%for MnSi,
but the value of $\lambda$ is not well known. For the upper and lower limits
used in Ref.\ \onlinecite{Belitz_Kirkpatrick_Rosch_2006b}, namely, $\lambda =
\epsilon_{\text{F}}/2$ and $\lambda = 540\,\text{K}$, respectively, we find
$T_{1-2} \approx 0.06\,T^*$ and $T_{1-2} \approx 500\,T^*$, which again
underscores the strong dependence on $\epsilon_{\text{F}}/\lambda$. Generally
speaking, weak helimagnets are the best candidates for observing the
weak-disorder transport regime.

Third, the anisotropic helimagnon dispersion relation, Eq.\ (\ref{eq:4a}),
leads to a 3-$d$ helimagnetic system being qualitatively similar to a 2-$d$ one
with just a Coulomb interaction. The components of the soft wave vector ${\bm
k}$ scale as $k_z \sim k_{\perp}^2 \sim T$. With $f({\bm k})$ a generic
function, the relevant integrals that determine observables are of the form
\begin{eqnarray*}
\int dk_z \int d{\bm k}_{\perp}\,{\bm k}_{\perp}^2\ \delta(\Omega^2 - k_z^2 -
{\bm k}_{\perp}^4)\,f(k_z,{\bm k}_{\perp}) &\propto&
\nonumber\\
&& \hskip -150pt \int d{\bm k}_{\perp}\,{\bm
k}_{\perp}^2\,\frac{\Theta(\Omega^2-{\bm k}_{\perp}^4)}{\sqrt{\Omega^2 - {\bm
k}_{\perp}^4}}\,f(k_z=0,{\bm k}_{\perp}),
\end{eqnarray*}
and the dependence of $f$ on $k_z$ can be dropped since it does not contribute
to the leading temperature scaling. The prefactor of the ${\bm k}$ dependence
of $f$ is of $O(1)$ in a scaling sense. As a result, the 3-$d$ ${\bm
k}$-integral behaves effectively like the integral in the 2-$d$ Coulomb case
\cite{2d_footnote}. In addition to the conductivity, these considerations apply
to weak-disorder corrections to the spin susceptibility, and to the specific
heat \cite{us_tbp}.

Finally, we point out a generic feature relevant for the explanation of the
$T^{3/2}$ dependence of the resistivity observed in the disordered phase of
MnSi. The prefactor of $T/\epsilon_{\text{F}}$ in
$\delta\sigma_{\text{L}}/\sigma_0$, Eq.\ (\ref{eq:10}), is very small. Assuming
$\nu = 1$ and the above parameter values for MnSi, it is about $3\times
10^{-5}$ for $\lambda = \epsilon_{\text{F}}/2$, and about $10^{-2}$ for
$\lambda = 540\,\text{K}$. The small prefactor reflects the long wavelength of
the helix on a microscopic scale. The prefactor of the $T^{5/2}$ behavior in
the clean limit is also small, for the same reason.
%This is consistent with the fact that so far no anomalous $T$-dependence of
%the resistivity has observed in the ordered phase of MnSi.
By contrast, the prefactor of the experimentally
observed $(T/\epsilon_{\text{F}})^{3/2}$ behavior of the resistivity in the
disordered phase is of $O(10^6)$. Therefore, the processes causing the
$T^{3/2}$ behavior in MnSi must occur on short length scales, and are unlikely
to be related to the helical order.

In conclusion, we predict that in the ordered phase of metallic helimagnets, at
low temperatures in the weak-disorder regime, the leading temperature
correction to the Drude conductivity is proportional to $T\tau$, and the size
of this effect depends on the direction of the current. This small but very
remarkable effect should be most easily observable in weak helimagnets.

This research was supported by the National Science Foundation under Grant Nos.
DMR-05-30314, DMR-05-29966, and PHY05-51164.

\vskip -1mm
%\bibliography{letter}

\begin{thebibliography}{13}
\expandafter\ifx\csname natexlab\endcsname\relax\def\natexlab#1{#1}\fi
\expandafter\ifx\csname bibnamefont\endcsname\relax
  \def\bibnamefont#1{#1}\fi
\expandafter\ifx\csname bibfnamefont\endcsname\relax
  \def\bibfnamefont#1{#1}\fi
\expandafter\ifx\csname citenamefont\endcsname\relax
  \def\citenamefont#1{#1}\fi
\expandafter\ifx\csname url\endcsname\relax
  \def\url#1{\texttt{#1}}\fi
\expandafter\ifx\csname urlprefix\endcsname\relax\def\urlprefix{URL }\fi
\providecommand{\bibinfo}[2]{#2} \providecommand{\eprint}[2][]{\url{#2}}

\bibitem[{\citenamefont{Ishikawa et~al.}(1976)\citenamefont{Ishikawa, Tajima,
  Bloch, and Roth}}]{Ishikawa_et_al_1976}
\bibinfo{author}{\bibfnamefont{Y.}~\bibnamefont{Ishikawa}},
  \bibinfo{author}{\bibfnamefont{K.}~\bibnamefont{Tajima}},
  \bibinfo{author}{\bibfnamefont{D.}~\bibnamefont{Bloch}}, \bibnamefont{and}
  \bibinfo{author}{\bibfnamefont{M.}~\bibnamefont{Roth}},
  \bibinfo{journal}{Solid State Commun.} \textbf{\bibinfo{volume}{19}},
  \bibinfo{pages}{525} (\bibinfo{year}{1976}).

\bibitem[{LRO()}]{LRO_footnote}
\bibinfo{note}{Strictly speaking, there never is any true long-range helical
  order at any nonzero temperature due to strong flucutations of the helix, see
  Ref.\ \onlinecite{Kirkpatrick_Belitz_2006}. However, this is a weak effect
  that we will neglect for our present purposes.}

\bibitem[{\citenamefont{Pfleiderer et~al.}(1997)\citenamefont{Pfleiderer,
  McMullan, Julian, and Lonzarich}}]{Pfleiderer_et_al_1997}
\bibinfo{author}{\bibfnamefont{C.}~\bibnamefont{Pfleiderer}},
  \bibinfo{author}{\bibfnamefont{G.~J.} \bibnamefont{McMullan}},
  \bibinfo{author}{\bibfnamefont{S.~R.} \bibnamefont{Julian}},
  \bibnamefont{and} \bibinfo{author}{\bibfnamefont{G.~G.}
  \bibnamefont{Lonzarich}}, \bibinfo{journal}{Phys. Rev. B}
  \textbf{\bibinfo{volume}{55}}, \bibinfo{pages}{8330} (\bibinfo{year}{1997}).

\bibitem[{\citenamefont{Pfleiderer et~al.}(2001)\citenamefont{Pfleiderer,
  Julian, and Lonzarich}}]{Pfleiderer_Julian_Lonzarich_2001}
\bibinfo{author}{\bibfnamefont{C.}~\bibnamefont{Pfleiderer}},
  \bibinfo{author}{\bibfnamefont{S.~R.} \bibnamefont{Julian}},
  \bibnamefont{and} \bibinfo{author}{\bibfnamefont{G.~G.}
  \bibnamefont{Lonzarich}}, \bibinfo{journal}{Nature}
  \textbf{\bibinfo{volume}{414}}, \bibinfo{pages}{427} (\bibinfo{year}{2001}).

\bibitem[{bal()}]{ballistic_footnote}
\bibinfo{note}{This regime is also referred to as the intermediate-temperature
  regime, or the ballistic regime. Here we refer to it as the weak-disorder
  regime, in order to avoid a connotation of ballistic transport in mesoscopic
  systems.}

\bibitem[{\citenamefont{Zala et~al.}(2001)\citenamefont{Zala, Narozhny, and
  Aleiner}}]{Zala_Narozhny_Aleiner_2001}
\bibinfo{author}{\bibfnamefont{G.}~\bibnamefont{Zala}},
  \bibinfo{author}{\bibfnamefont{B.~N.} \bibnamefont{Narozhny}},
  \bibnamefont{and} \bibinfo{author}{\bibfnamefont{I.~L.}
  \bibnamefont{Aleiner}}, \bibinfo{journal}{Phys. Rev. B}
  \textbf{\bibinfo{volume}{64}}, \bibinfo{pages}{214204}
  (\bibinfo{year}{2001}).

\bibitem[{\citenamefont{Sergeev et~al.}(2005)\citenamefont{Sergeev, Reizer, and
  Mitin}}]{Sergeev_Reizer_Mitin_2005}
\bibinfo{author}{\bibfnamefont{A.}~\bibnamefont{Sergeev}},
  \bibinfo{author}{\bibfnamefont{M.~Y.} \bibnamefont{Reizer}},
  \bibnamefont{and} \bibinfo{author}{\bibfnamefont{V.}~\bibnamefont{Mitin}},
  \bibinfo{journal}{Phys. Rev. B} \textbf{\bibinfo{volume}{69}},
  \bibinfo{pages}{075310} (\bibinfo{year}{2005}).

\bibitem[{pin()}]{pinning_footnote}
\bibinfo{note}{In a rotationally invariant model, effects of helical
  fluctuations will be strongest in the ordered phase. In a crystal,
  crystal-field effects will pin the helix, which makes the helical excitations
  massive. Such pinning is expected to disappear in a phase where the helical
  order is destroyed, but still persists on some length scales. A rotationally
  invariant model of the ordered phase may therefore provide a reasonable
  description of the high-pressure phase, and insight into the transport
  properties of the ordered phase might give clues about the origin of the
  observed $T^{3/2}$ behavior.}

\bibitem[{\citenamefont{Belitz et~al.}(2006)\citenamefont{Belitz, Kirkpatrick,
  and Rosch}}]{Belitz_Kirkpatrick_Rosch_2006b}
\bibinfo{author}{\bibfnamefont{D.}~\bibnamefont{Belitz}},
  \bibinfo{author}{\bibfnamefont{T.~R.} \bibnamefont{Kirkpatrick}},
  \bibnamefont{and} \bibinfo{author}{\bibfnamefont{A.}~\bibnamefont{Rosch}},
  \bibinfo{journal}{Phys. Rev. B} \textbf{\bibinfo{volume}{74}},
  \bibinfo{pages}{024409} (\bibinfo{year}{2006}).

\bibitem[{dia()}]{diagonality_footnote}
\bibinfo{note}{It is not obvious that it is possible to find a quasi-particle
  description where the Green function is diagonal in momentum space. However,
  the canonical transformation mentioned in the text shows that it is.}

\bibitem[{us_()}]{us_tbp}
\bibinfo{note}{T.~R. Kirkpatrick, D. Belitz, and Ronojoy Saha, unpublished
  results.}

\bibitem[{2d_()}]{2d_footnote}
\bibinfo{note}{It is important for this argument that there is no qualitative
  difference between a stress coupling and a density coupling. This is true in
  the weak-disorder limit but not, for instance, in the diffusive limit.}

\bibitem[{\citenamefont{Kirkpatrick and
  Belitz}(2006)}]{Kirkpatrick_Belitz_2006}
\bibinfo{author}{\bibfnamefont{T.~R.} \bibnamefont{Kirkpatrick}}
  \bibnamefont{and} \bibinfo{author}{\bibfnamefont{D.}~\bibnamefont{Belitz}},
  \bibinfo{journal}{Phys. Rev. Lett.} \textbf{\bibinfo{volume}{97}},
  \bibinfo{pages}{267205} (\bibinfo{year}{2006}).

\end{thebibliography}

\end{document}